\documentclass[aps,a4paper,showpacs,amsfonts,amsmath,amssymb,superscriptaddress]{revtex4}



\usepackage{graphicx}
\usepackage{dcolumn}

\begin{document}

\title{Scaling laws of strategic behaviour and size heterogeneity in agent dynamics}

\author{Gabriella Vaglica}
\affiliation{Dipartimento di Fisica e Tecnologie Relative, Universit\`a di Palermo,
Viale delle Scienze, I-90128, Palermo, Italy}

\author{Fabrizio Lillo}
\affiliation{Dipartimento di Fisica e Tecnologie Relative, Universit\`a di Palermo,
Viale delle Scienze, I-90128, Palermo, Italy}
\affiliation{Santa Fe Institute, 1399 Hyde Park
Road, Santa Fe, NM 87501, USA}

\author{Esteban Moro}
\affiliation{Grupo Interdisciplinar de Sistemas Complejos (GISC)\\Departamento de Matem\'aticas, Universidad Carlos III de Madrid, Avedida de la Universidad 30, E-28911, Legan\'es, Spain}

\author{Rosario N. Mantegna}
\affiliation{Dipartimento di Fisica e Tecnologie Relative, Universit\`a di Palermo,
Viale delle Scienze, I-90128, Palermo, Italy}

\date{\today}

\begin{abstract}

The dynamics of many socioeconomic systems is determined by the decision making process of agents. The decision process depends on agent's characteristics, such as preferences, risk aversion, behavioral biases, etc. \cite{Kahneman1979,Lux1999}. In addition, in some systems the size of agents can be highly heterogeneous leading to very different impacts of agents on the system dynamics \cite{Pareto1897, Zipf1949, Ijiri1977, Axtell2001, Pushkin04, Gabaix06}. The large size of some agents poses challenging problems to agents who want to control their impact, either by forcing the system in a given direction or by hiding their intentionality. Here we consider the financial market as a model system, and we study empirically how agents strategically adjust the properties of large orders in order to meet their preference and minimize their impact. We quantify this strategic behavior by detecting scaling relations of allometric nature \cite{Calder1984} between the variables characterizing the trading activity of different institutions.
We observe power law distributions in the investment time horizon, in the number of transactions needed to execute a large order and in the traded value exchanged by large institutions and we show that heterogeneity of agents is a key ingredient for the emergence of some aggregate properties characterizing this complex system.

\end{abstract}

\maketitle

In many complex systems agents self organize themselves in an ecology of different  ``species"  interacting in a variety of ways. 
Agents are not only different in their strategies, information, and preferences, but they can be very different in their size. Examples include individual's wealth \cite{Pareto1897} and firms size \cite{Ijiri1977,Axtell2001}. The presence of agents with large size poses several challenging questions. It is likely that large agents impacts the system in a way that is significantly different from small ones. Indeed, small agents can easily hide their intentionality, while for large agents this is not so easy and they must adopt strategies taking into account their own effect because revealing their intention could decrease their fitness.

Financial markets are an ideal system to investigate this problem. There is empirical evidence that market participants are very heterogeneous in size. For example banks \cite{Pushkin04} and mutual funds \cite{Gabaix06} size follow Zipf's law, i.e. the probability that the size of a participant is larger than $x$ decays as $1/x$ \cite{Zipf1949}. As a consequence large investors usually need to trade large quantities that  can significantly affect  prices. The associated cost is called market impact \cite{Hasbrouck1991,Hausman1992,Dufour2000,Plerou2001,Lillo2003,Bouchaud2004}. For this reason large investors refrain from revealing their demand or supply and they typically trade their large orders incrementally over an extended period of time. These large orders are called {\it packages} \cite{Chan95,Gallagher2006} or {\it hidden orders} and are split in smaller trades as the result of a complex optimization procedure which takes into account the investor's preference, risk aversion, investment horizon, etc.. 

Here we investigate the trading activity of a large fraction of the financial firms exchanging a financial asset at the Spanish Stock Market (Bolsas y Mercados Espa\~noles, BME) in the period 2001-2004 (see Materials and Methods section for a description of data). Firms are credit entities and investment firms which are members of the stock exchange and are entitled to trade in the market. 

Our approach aims to be a comprehensive approach analysing the overall dynamics of all packages exchanged in the market. However, our database does not contain direct information on packages, so that this information must be statistically inferred from the available data. 
Since we do not have information on clients but only on firms, we develop a detection algorithm (see Material and Methods for a description of the algorithm) which is not sensible to small fluctuation in the buy/sell activity of a firm. The algorithm detects time segments in the inventory time evolution of a firm when the firm acts as a net buyer or seller at an approximately constant rate. We call these segments {\it patches} and we assume that in each of these patches it is contained at least one package.

\begin{figure}
\includegraphics[scale=0.3,angle=-90]{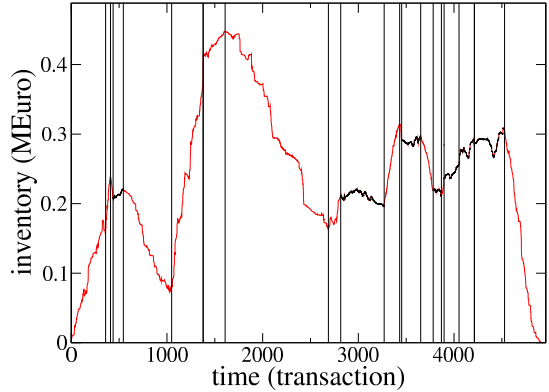}
\caption{Example of an inventory time series. The series refers to a particular firm trading Santander. The vertical lines indicate the position where our algorithm predicts the boundary between two patches. The red patches are directional patches. Due to their statistical nature, in each patch there are buy (with a total traded value $V_b$) and sell (with a total traded value $V_s$) trades. We consider directional patches, i.e. patches where either $V_b/V>\theta$ (buy patch) or $V_s/V>\theta$ (sell patch), where $V=V_b+V_s$. For buy patches $V_m=V_b$ whereas for sell patches $V_m=V_s$. In the present study we set  $\theta=75\%$. Directional patches are shown as red lines. The black patches are not directional and are not considered in the rest of the paper.}
\label{series} 
\end{figure}

Since firms act simultaneously as brokers for many clients, it is rather frequent that in a patch not all the transactions have the same sign. However, a vast majority of firm inventory time series can be partitioned in patches with a well defined direction to buy or to sell. This is probably due to the fact that in most cases the trading activity of a firm is dominated by the activity of one big client. 
We consider {\it directional patches}, i.e. patches with a well defined direction (see Figure~\ref{series}).
The characterizing variables of a directional patch are the time length $T$ (in seconds) of the patch, measured as the time interval between the first and the last order of the patch, the traded value $V_m$ and the number $N_m$ of trades characterizing the patch. For example, $N_m$ is the number of buy trades and $V_m$ is the purchased value for buy patches.

\begin{figure}
\includegraphics[scale=0.25,angle=-90]{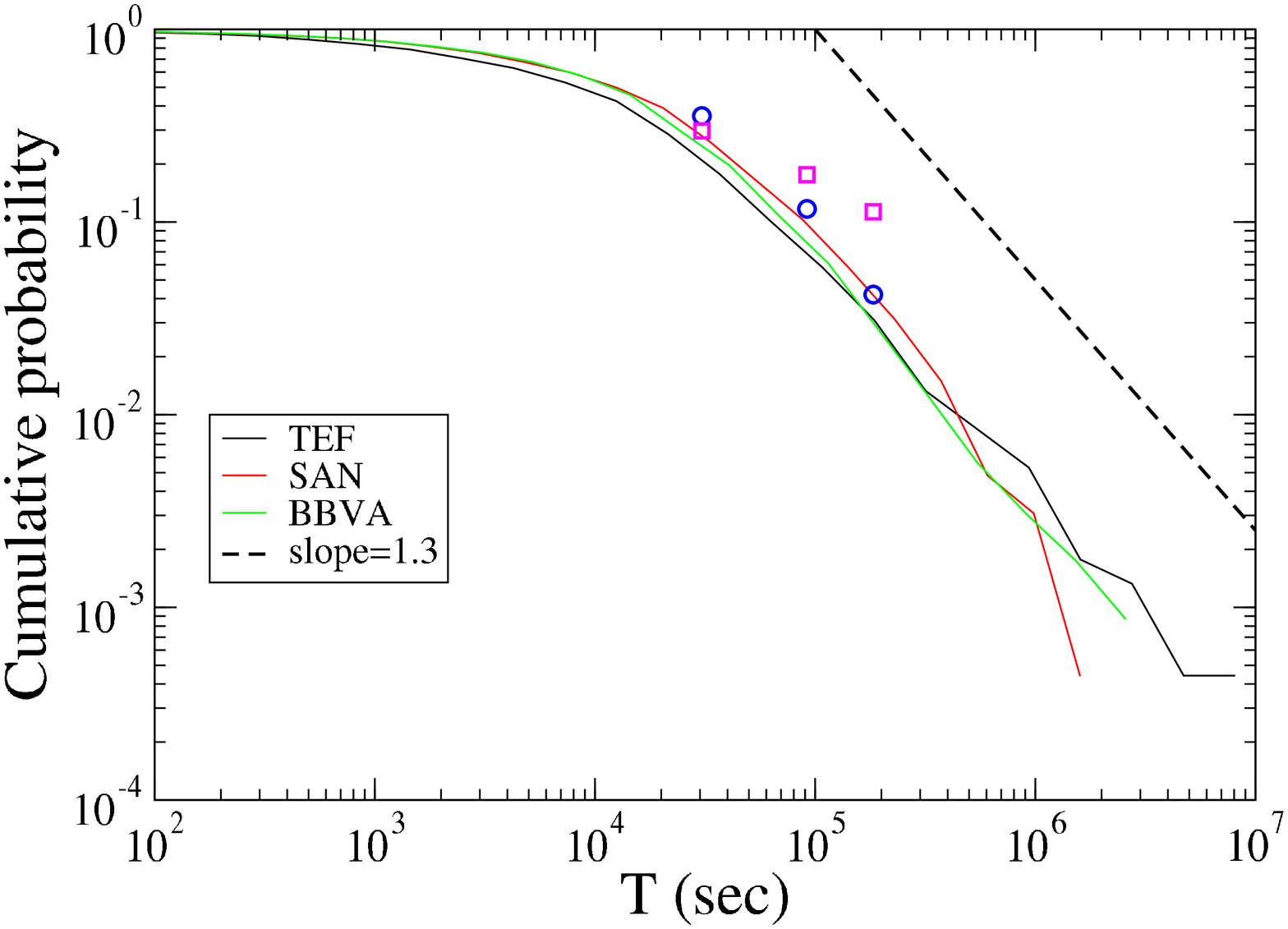}
\includegraphics[scale=0.25,angle=-90]{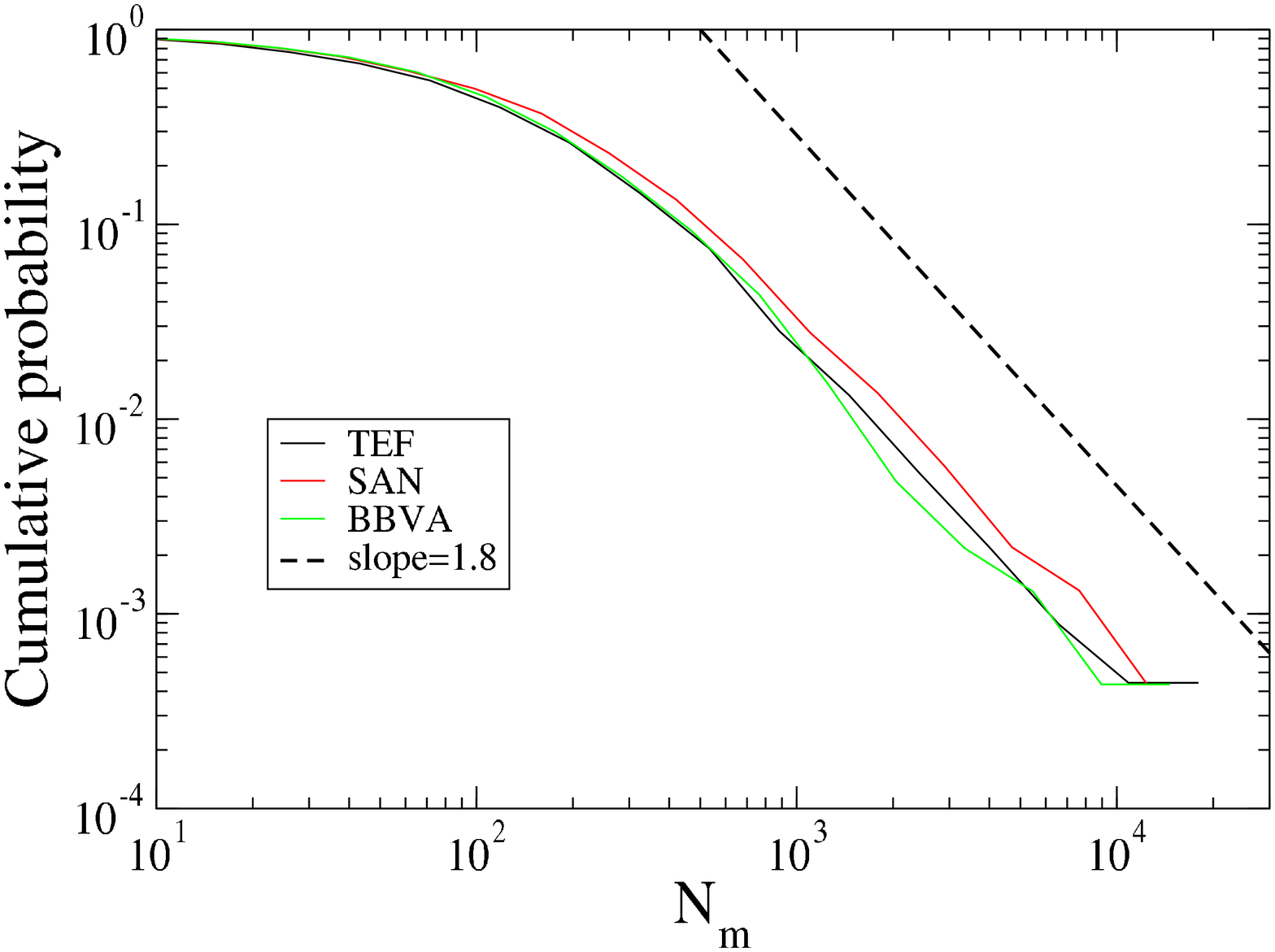}
\includegraphics[scale=0.25,angle=-90]{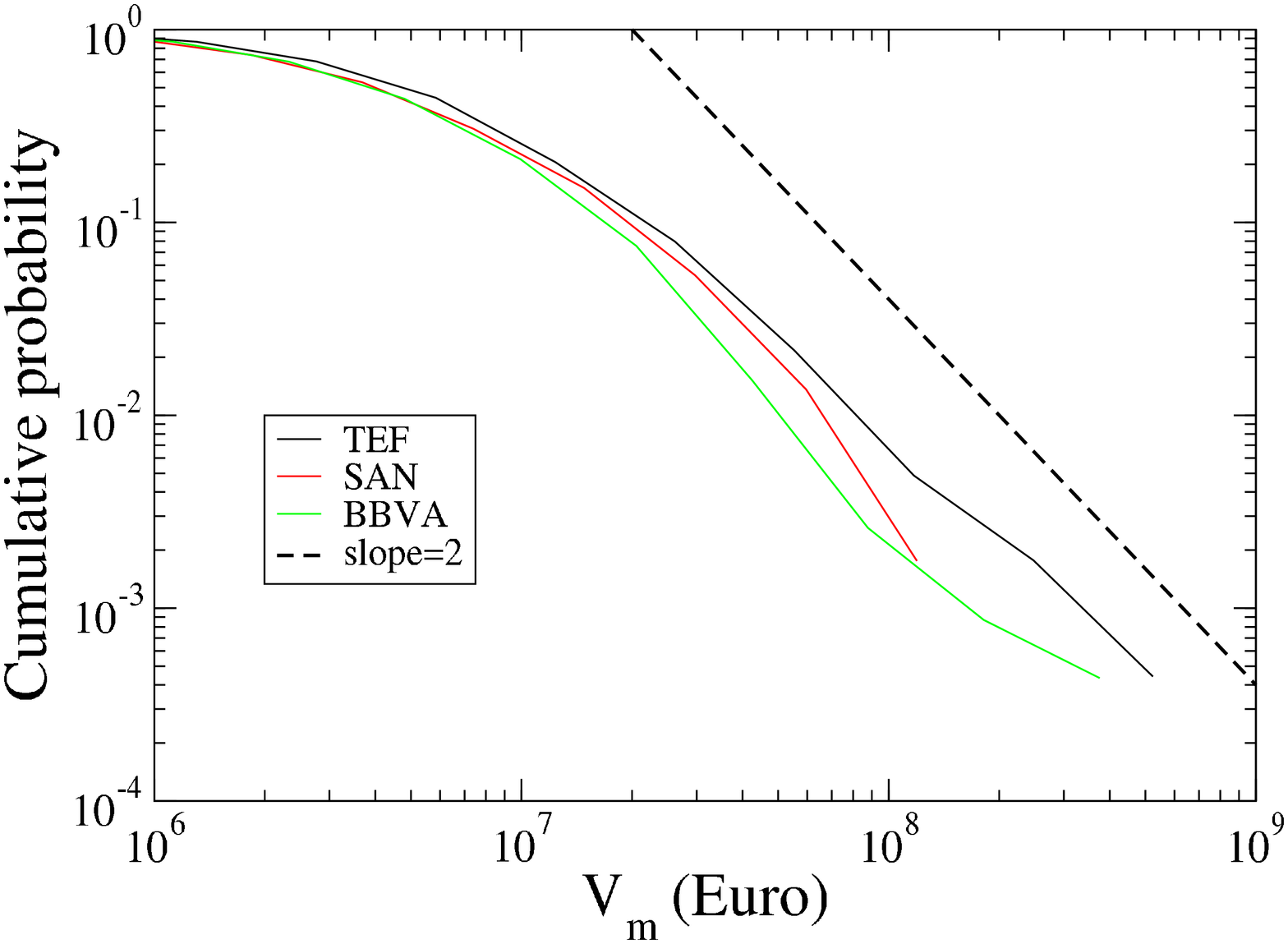}
\caption{Distribution of $T$, $N_m$, and $V_m$ for the the stocks  Banco Bilbao Vizcaya Argentaria (BBVA), Banco Santander Central Hispano (SAN), and Telef\'onica (TEF). In the panel showing the distribution of $T$ we plot the distribution of packages reported in literature on packages. Specifically, blue circles are results from Ref.~\cite{Chan95} for packages traded at the New York Stock Exchange and magenta squares are results from Ref.~\cite{Gallagher2006} for packages traded at the Australian Stock Exchange. }
\label{distrib} 
\end{figure}

We investigate first the distributional properties of the patches identified by our algorithm.  Figure ~\ref{distrib} shows the distribution of $T$, $N_m$, and $V_m$  for the three investigated stocks. The asymptotic behavior of all the three distributions can be approximated by a power law function $P(X)\sim 1/X^{\zeta_X+1}$, where $X$ can be $T$, $N_m$, or $V_m$ and $\zeta_X$ is the exponent characterizing the power law behavior. A summary of the estimated exponents is shown in Table~\ref{summary} from which one can conclude that $\zeta_{V_m}\simeq 2$, $\zeta_{N_m}\simeq 1.8$, and $\zeta_T\simeq 1.3$. Our analysis makes explicit the presence of very broad distribution for the three variables characterizing a patch. In fact the very low value of the exponents is consistent with the conclusion that $T$ and $N_m$ belong to the domain of L\'evy stable distributions. This result indicates that in the market there is a huge heterogeneity in the scales characterizing the trading profile of the investors. 

The  volume of the packages is likely to be related to the size of the investors. Large investors need to trade large packages to rebalance their portfolio. Gabaix {\it et al.} \cite{Gabaix2003} developed a theory which predicts that package size should be power law distributed with an exponent $\zeta_{V_m}=3/2$. The value we find for $\zeta_{V_m} \simeq 2.$ is slightly larger than the one predicted by them. 
On the contrary, the value $\zeta_{N_m}=3$ derived by the theory in \cite{Gabaix2003} is significantly larger than our estimate ($\zeta_{N_m} \simeq 1.8$). 
Finally, the power law distribution of packages time length $T$ might reflect the heterogeneity of time scales among investors. 
The distribution of $T$ is compatible with the ones obtained by using specialized database describing the investment packages of large investors \cite{Chan95,Gallagher2006} (see Figure~\ref{distrib}). 
Gabaix {\it et al.} theory \cite{Gabaix2003} predicts the value $\zeta_T=3$ which is significantly larger than our value ($\zeta_T \simeq 1.3$). The presence of power law distribution of investors time scales has been recently suggested in stylized models of investment decisions \cite{Borland,Lillo07,Eisler}.

\begin{table}
\caption{Summary of the properties of detected patches. The number in parenthesis nearby the tick symbol is the number of patches detected for the considered stock. Rows 1-3: Tail exponents of the distribution of $T$, $N_m$, and $V_m$ estimated with the Hill estimator (or Maximum  Likelihood Estimator). In parenthesis we report the $95\%$ confidence interval. Rows 4-6: Exponents of the allometric relations defined in Eq.~\ref{scaling}. The exponents are estimated with PCA and the errors are estimated with bootstrap.  In parenthesis we report the $95\%$ confidence interval. Rows 7-9: Percentage of firms with at least $10$ patches for which one cannot reject the hypothesis of lognormality with $95\%$ confidence according to Jarque-Bera test. The numbers in parenthesis are the number of firms for which one cannot reject the hypothesis of lognormality divided to the number of firms used in the test. 
}
\begin{tabular}{r||c|c|c|}
&BBVA~~~(2104)&SAN~~~(2086)&TEF~~~(2062)\\
 \hline
$\zeta_{V_m}$&$2.3~~~(1.9~;2.7)~~~$&$2.0~~~(1.7~;2.3)~~~$&$1.9~~~(1.6~;2.2)~~~$\\
$\zeta_{N_m}$&$2.0~~~(1.7~;2.3)~~~$&$1.7~~~(1.4~;2.0)~~~$&$1.7~~~(1.4~;2.0)~~~$\\ 
$\zeta_T$&$1.5~~~(1.3~;1.7)~~~$&$1.5~~~(1.3~;1.7)~~~$&$1.2~~~(1.0~;1.4)~~~$\\
\hline
$g_1$&$1.08~~~(1.05~;1.12)~~~$&$1.06~~~(1.01~;1.10)~~~$&$1.07~~~(1.04~;1.11)~~~$\\
$g_2$&$1.81~~~(1.69~;1.93)~~~$&$1.81~~~(1.68~;1.94)~~~$&$2.00~~~(1.88~;2.14)~~~$\\
$g_3$&$0.68~~~(0.65~;0.71)~~~$&$0.68~~~(0.65~;0.70)~~~$&$0.62~~~(0.59~;0.64)~~~$\\
\hline
$T$&$75$ $(15/20)$&$63$ $(17/27)$&$77$ $(24/31)$\\ 
$N_m$&$90$ $(18/20)$&$100$ $(27/27)$&$100$ $(31/31)$\\
$V_m$&$90$ $(18/20)$&$100$ $(27/27)$&$94$ $(29/31)$\\
\end{tabular}
\label{summary}
\end{table}

\begin{figure}
\includegraphics[scale=0.25,angle=-90]{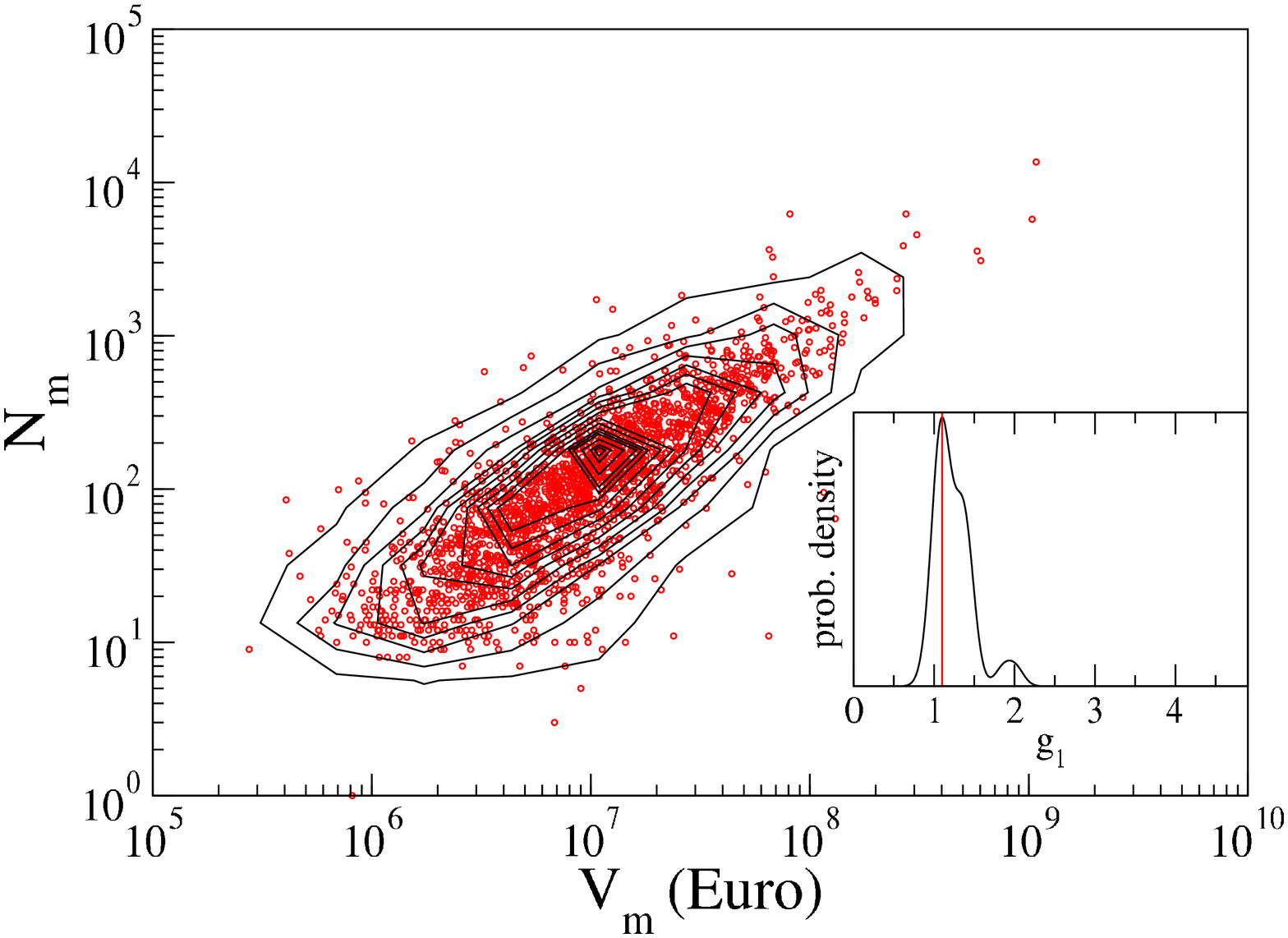}
\includegraphics[scale=0.25,angle=-90]{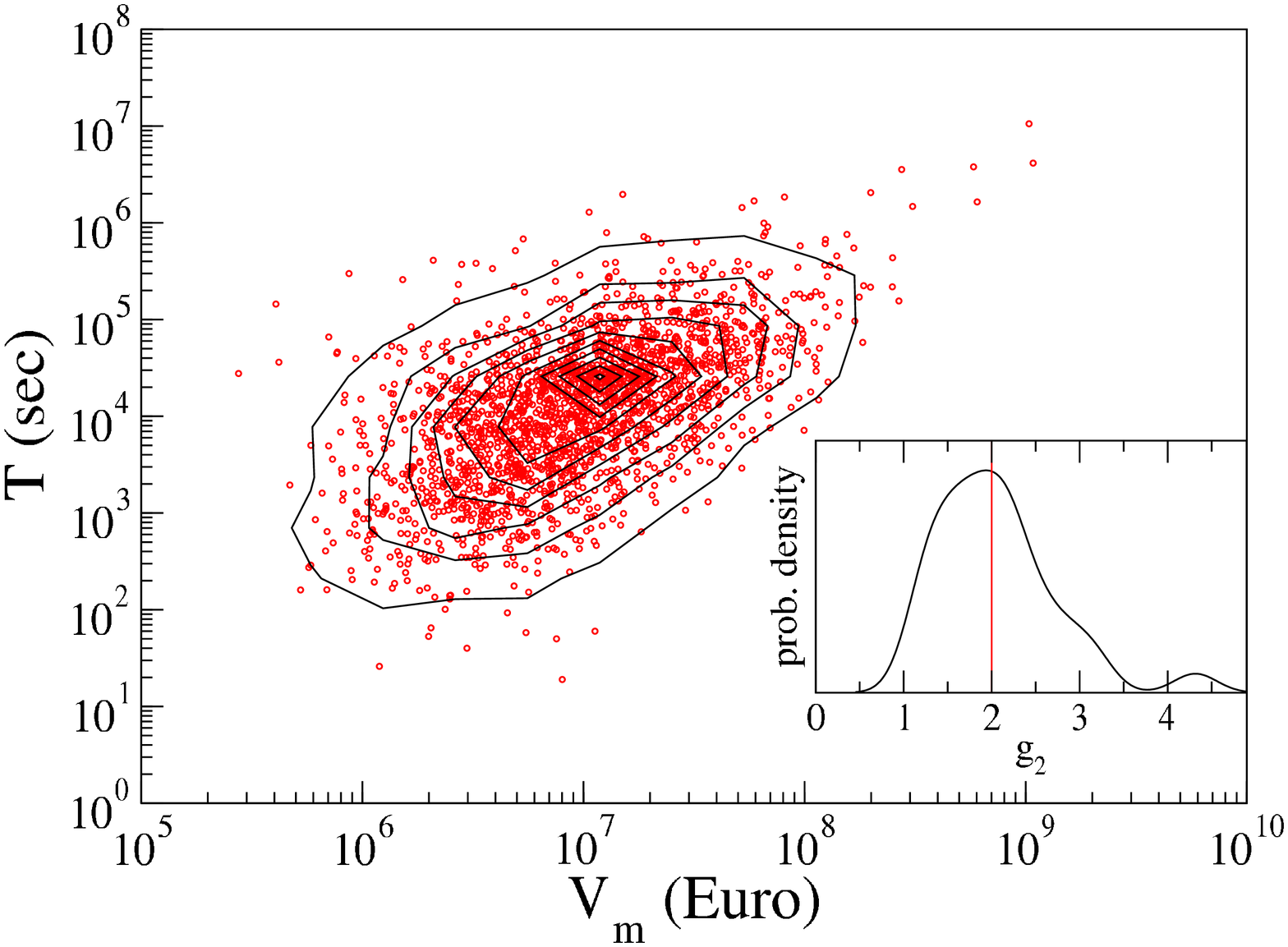}
\includegraphics[scale=0.25,angle=-90]{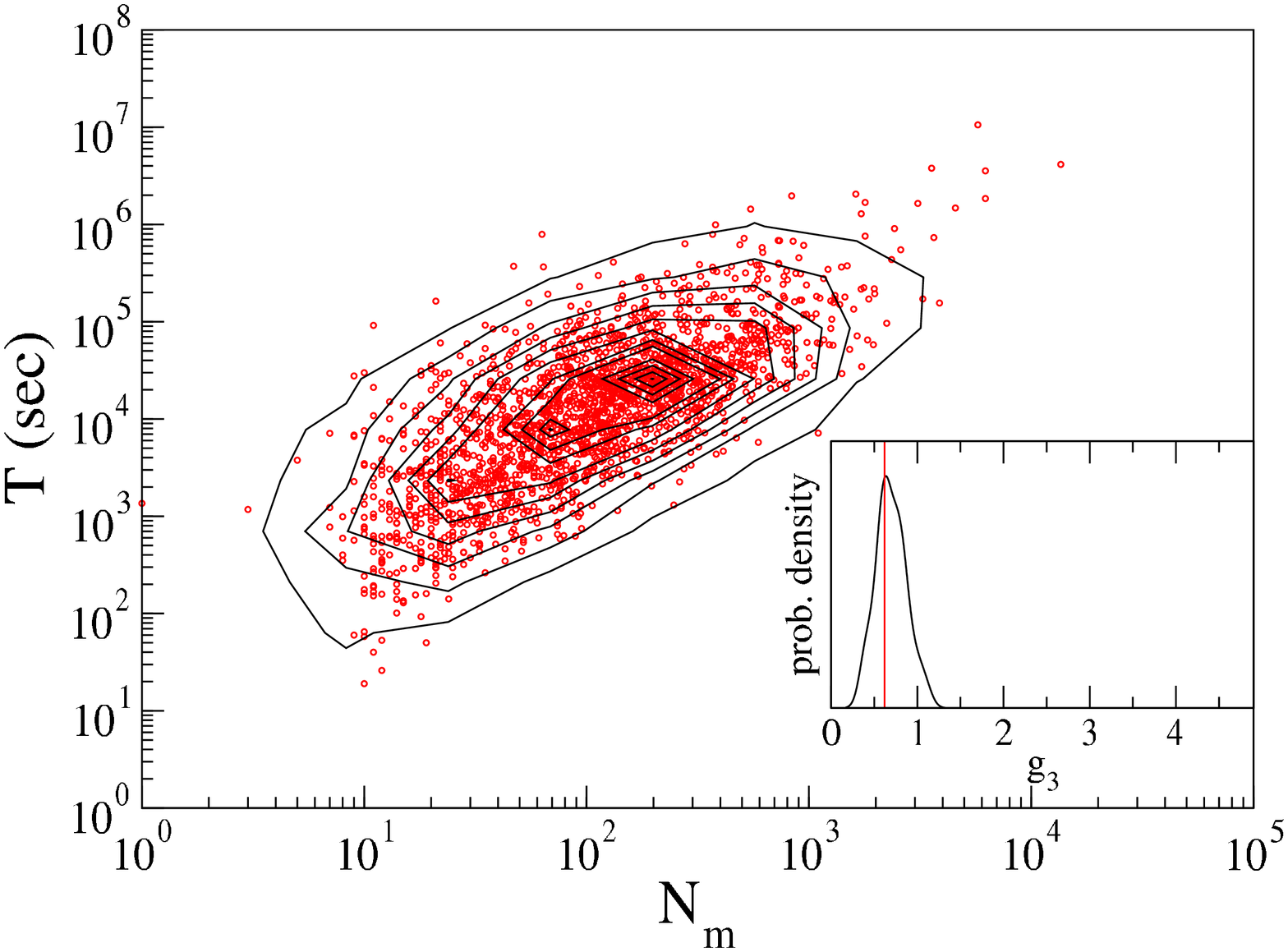}
\caption{Scatter plots of the variables $T$, $N_m$, and $V_m$ for Telef\'onica. The black lines are contour lines of the bivariate probability density function. The insets show the probability density functions of the three exponents $g_1$, $g_2$, and $g_3$ describing the allometric relations of Eq.~\ref{scaling} computed on the patches of individual firms with at least $10$ patches.  The red vertical lines indicates the values of the scaling exponents computed in the pool of all firms and reported in rows 4-6 of Table~\ref{summary}.  It is worth noting that the dispersion of $g_2$ is significantly larger than the one for the other two exponents.}
\label{scatter} 
\end{figure}

The role of size heterogeneity in the emergence of power law distributions will be considered at the end of the paper. To complete our characterization of firm patches, we now consider the relation between the variables characterizing each patch. Specifically, 
by applying the Principal Component Analysis (PCA) to the set of points with coordinates $(\log T,\log N_m,\log V_m)$, we investigate the allometric relations between any two of the above variables, i.e.
\begin{equation}
N_m\sim V_m^{g_1}~~~~~~T\sim V_m^{g_2}~~~~~~~N_m\sim T^{g_3}
\label{scaling}
\end{equation}
Figure~\ref{scatter}  shows the scatter plots and the contour plots for the stock Telef\'onica. In all three cases a clear dependence between the variables is seen.  PCA analysis shows that the first eigenvalue explains on average $91\%$, $83\%$, and $89\%$ of the variance for the first, second, and third allometric relation, respectively, indicating a strong correlation between the variables. The estimated exponents (see Table~\ref{summary}) are consistent for different stocks so that the allometric relations are
\begin{equation}
N_m\sim V_m^{1.1}~~~~~~T\sim V_m^{1.9}~~~~~~~N_m\sim T^{0.66}
\label{scaling2}
\end{equation}
The presence of scaling relations between the variables were first suggested in Ref.~\cite{Gabaix2003} but
it is worth noting that the theory developed in that paper predicts $g_1=g_2=1/2$ and $g_3=1$, and these values are quite different from the ones we estimate from data.
The first allometric relation indicates that the number of transactions in which a package is split is approximately proportional to the total traded value of the package. This implies that the mean transaction volume is roughly independent on the size of the package.  
This mean value is on average determined by the size of the available volume at the best quote indicating that the trader does not trade orders larger than the volume available at the best quote, probably to avoid being  too aggressive \cite{Farmer04}.

We consider the relation between the three variables together by performing a PCA on the set of points describing the patches and identified by the coordinates $(\log T,\log N_m,\log V_m)$ \cite{Sprent1972}. 
The set of points effectively lies on a two dimensional manifold which has one dimension much larger than the other. The fact that the first eigenvalue is large indicates that one factor dominates the trading strategy. 
The allometric relations of the three variables associated with the first eigenvalue of the PCA provides an estimation of the exponents ($g_1\simeq 1.2$, $g_2\simeq 1.8$, and $g_3\simeq 0.67$ for Telef\'onica) which, differently than in the bivariate case, are of course coherent among them and only slightly different from the ones obtained from the bivariate analysis. 

We now go back to the problem of assessing the role of firm heterogeneity. The first scientific question is: Is the fat tailed distribution of $T$, $N_m$, and $V_m$  due to the fact that individual firms place heterogeneously sized packages or is this an effect of the aggregation of many different firms together? To answer this question we test the hypothesis that the patches identified for a given firm trading a given stock are lognormally distributed. The test (see Table~\ref{summary}) shows that for most of the trading firms we cannot reject the hypothesis that the patches have characteristics sizes distributed as a lognormal. Since we reject the lognormal hypothesis for the pool obtained by considering all the firms, we conclude that the power law distribution of  $T$, $N_m$ and $V_m$ is  due to an heterogeneity in patch scale {\it between} different firms rather that {\it within} each firm. 
The second scientific question about concerns the role of firm heterogeneity for scaling laws. To assess the role of heterogeneity, for each firm we compute the exponents $g_1$, $g_2$, and $g_3$ of the bivariate relations of Eq. \ref{scaling} (see insets of fig.~\ref{scatter}). We observe that the exponents obtained for each firm are distributed around the corresponding value of the exponent obtained for the pool. This result indicates that the bivariate allometric relations are not an effect of the aggregation but are observed, on average, also for individual firms.

In conclusion our comprehensive investigation of packages traded at BME shows that heterogeneity of firms has an essential role for the emergence of power law tails in the investment time horizon, in the number of transactions and in the traded value exchanged by packages. Differently, scaling laws between the variables characterizing each package are essentially the same across different firms with the possible exception of the relation between $T$ and $V_m$ perhaps reflecting different degree of aggressiveness of firms.


\section{MATERIALS AND METHODS}

Our database of the electronic open market SIBE (Sistema de Interconexi\'on Burs\'atil Electr\'onico) allows us to follow each transaction performed by all the firms registered at BME.  In 2004 the BME was the eight in the world in market capitalization.
We consider the stocks  Banco Bilbao Vizcaya Argentaria (BBVA), Banco Santander Central Hispano (SAN), and Telef\'onica (TEF). We also consider only the most active firms defined by the criterion that each firm made at least $1,000$ trades/year and was active at least $200$ days per year.  The number of firms is $50$ (BBVA), $55$ (SAN), and $61$ (TEF). These firms are involved in $81-86\%$ of the transactions.  The investigated period is 2001-2004. We do not consider other stocks because we have verified that the number of detected patches is too small to perform careful statistical estimation.

The series under study is the series of signed traded value. For each firm and for each stock we construct the series composed by all the trades performed by the firm with a value $+v$ for a buy trade and $-v$ for a sell trade, where $v$ is the value (in Euros) of the traded shares. 

The method we use to detect statistically the presence of patches is adapted from Ref.~\cite{Bernaola01} where it was introduced to study patchiness non-stationarity of human heart rate. The algorithm works as follows. One moves a sliding pointer along the signal and computes the mean of the subset of the signal to the left and to the right of the pointer. From these mean values one computes a $t$ statistics and finds the position of the pointer for which the $t$ statistics is maximal. The significance level of this value of $t$ is defined as the probability of obtaining it or a smaller value in a random sequence. One then chooses a threshold (in our case $99\%$) and the sequence is cut if the significance level is smaller than the threshold. The cut position is the boundary between two consecutive patches. The procedure continues recursively on the left and right subset created by each cut. Before a new cut is accepted one also computes  $t$ between the right-hand new segment and its right neighbor and $t$ between the left-hand new segment and its left neighbor and one checks if both values of $t$ are statistically significant according to the selected threshold. The process stops when it is not possible to make new cut with the selected significance.

In the present study, we are mainly interested in directional patches, i.e. patches where the trader consistently buys or sells a large amount of shares. In other words we wish to exclude patches in which the inventory of the firm is diffusing randomly, without a drift. To this end for each patch we compute the total value purchased $V_b$, the total value sold $V_s$ and the total value $V=V_b+V_s$.  We then consider a patch as directional when either $V_b/V>\theta$ (buy patch) or $V_s/V>\theta$ (sell patch). The parameter $\theta$ can be varied and in the present study we set it to $\theta=75\%$. We obtain similar results for different values of $\theta$ such as $85\%$ and $95\%$. Finally in the present paper we consider patches with at least $10$ trades.

{\bf Acknowledgments} Authors acknowledge Sociedad de Bolsas for providing the data and the Integrated Action Italy-Spain ``Mesoscopics of a stock market" for financial support. 
GV, FL, and RNM acknowledge support from MIUR research project ``Dinamica di altissima frequenza nei mercati finanziari'' and NEST-DYSONET 12911 EU project. EM acknowledges partial support from MEC (Spain) throught grants FIS2004-01001, MOSAICO and a Ram\'on y Cajal contract and Comunidad de Madrid through grants UC3M-FI-05-077 and SIMUMAT-CM

\end{document}